\documentclass[12pt,a4paper]{article}
\usepackage[latin1]{inputenc}

\usepackage{amsmath}
\usepackage{amsfonts}
\usepackage{amssymb}
\usepackage{graphicx}
\usepackage{float}
\usepackage{cite}
\usepackage{hyperref}
\numberwithin{equation}{section}

\newcommand{\be}{\begin{equation}}
\newcommand{\ee}{\end{equation}}
\newcommand{\p}{\partial}
\newcommand{\CA}{\mathcal{A}}
\newcommand{\CL}{\mathcal{L}}
\newcommand{\M}{\mathcal{M}}
\newcommand{\N}{\mathcal{N}}
\newcommand{\K}{\mathcal{K}}
\newcommand{\A}{\mathcal{A}}
\newcommand{\F}{\mathcal{F}}

\newcommand{\Tr}{\,\mathrm{Tr}\,}
\begin{document}
\begin{titlepage}
\thispagestyle{empty}

\begin{center}
\noindent{\Large \textbf{Quantum Local Quench, AdS/BCFT and Yo-Yo String
}}\\
\vspace{1cm}

%authors
\vspace{1cm}
Amin Faraji Astaneh$^{\rm a,b,}$\footnote{faraji@ipm.ir} and Amir Esmaeil Mosaffa $^{\rm a,}$\footnote{mosaffa@theory.ipm.ac.ir}

 \vspace*{0.25cm}
 \begin{quote}

\center $^{\rm a}$\,\, {\sl School of Particles and Accelerators,\\ Institute for Research in Fundamental Sciences (IPM), \\
 P.O. Box 19395-5531, Tehran, Iran}

\center $^{\rm b}$\,\, {\sl Department of Physics, Sharif University of Technology,\\
P.O. Box 11365-9161, Tehran, Iran}
 \end{quote}

\end{center}
\vspace{2cm}
\begin{abstract}
We propose a holographic model for local quench in $1+1$ dimensional
Conformal Field Theory (CFT). The local quench is produced by joining two
identical CFT's on semi-infinite lines. When these theories
have a zero boundary entropy, we use the AdS/Boundary CFT proposal
to describe this process in terms of bulk physics. Boundaries of the original
CFT's are extended in AdS as dynamical surfaces. In our holographic picture these
surfaces detach from the boundary and form a closed folded string which can propagate
in the bulk. The dynamics of this string
is governed by the tensionless Yo-Yo string solution and its subsequent evolution
determines the time dependence after quench. We use this model to calculate holographic
Entanglement Entropy (EE) of an interval as a function of time. We propose
how the falling string deforms Ryu-Takayanagi's curves. Using the deformed curves we calculate
EE
and find complete agreement
with field theory results.

\end{abstract}

\end{titlepage}
%%%%%%%%%%%%%%%%%%%%%%%%%%%%%%%%%%%%%%%%%%%%%%%%%%%%%%%%%%%%
\tableofcontents
%%%%%%%%%%%%%%%%%%%%%%%%%%%%%%%%%%%%%%%%%%%%%%%%%%%%%%%%%%%%
\section{Introduction}
Time dependent processes are of utmost interest in physics and most
of the time difficult to address. In this work we consider such a
process known as {\it{ Quantum Quench}} \cite{cc0711},\cite{Calabrese:2007rg} and  \cite{Calabrese:1002}. Our objective is to study
this phenomenon by means of holography\cite{Maldacena:1997re}, \cite{Witten:1998qj} and \cite{Aharony}.

The field theoretic setup is the one considered in \cite{Calabrese:2006rx};
two identical $1+1$ dimensional Conformal Field
Theories (CFT's) each living on a half line, and prepared in their
ground states, are joined together at an instant of time to produce
a CFT on the whole line. This process is called {\it{Local Quench}}
and the question of interest is the time evolution of the system
after quench \cite{2007JSMTE..06....5E}, \cite{2008JSMTE..01..023E} and \cite{2011JSMTE..10..027D}.

A natural probe to investigate the time evolution is
the Entanglement Entropy (EE) of various subspaces \cite{Calabrese:2006rx}-\cite{Albash:2010mv}. On the field theory
side, this problem has been addressed in full detail and rigour in
\cite{Calabrese:2006rx},\cite{Calabrese:2009qy} using the power of conformal symmetry.

In this work we address the same problem using holography. The system
we consider consists of two Boundary Conformal Field Theories (BCFT's)
on half line, each of which can be described holographically via the AdS/BCFT
correspondence of \cite{Takayanagi:2011zk}, see also \cite{Fujita:2011fp}.
In this setup, each BCFT is described by part of
the AdS$_3$ background which is bounded by the half plane of the BCFT on the one hand
and a co-dimension one hyper-surface (two dimensional in this case)
in the bulk on the other. The hyper-surface which
is a dynamical object is part of the holographic description and intersects the boundary
on a line.

In the holographic model that we propose for local quench, when the
two BCFT's are joined together, the corresponding hyper-planes detach
from the boundary and attach one another in the bulk.
This leaves the boundary of the whole system as the two dimensional world volume
of a CFT on a plane. The bulk will be the whole AdS$_3$ with
a two dimensional dynamical hyper-plane floating in it.
The time evolution of this hyper-plane in the bulk determines
that of the resulting CFT in the boundary.

The problem simplifies significantly if the
BCFT's have a zero value of the so called "boundary entropy" and this is
the problem we consider. The general case of nonzero boundary entropy is
postponed to future works.

As we will describe in subsequent sections, when the boundary entropy is zero,
the hyper-planes will become the world sheets of
tensionless open strings in the bulk with the usual Polyakov action.
 The strings attach the boundary
with a Neumann boundary condition. The quench will then correspond to
detaching the strings from the
boundary and attaching the tips together in the bulk.
This will produce a folded closed string in AdS
whose dynamics is determined by the Polyakov action.
The solution will turn out to be the
tensionless limit of the famous "Yo-Yo String" \cite{Bardeen:1975gx},
\cite{Ficnar:2013wba}. The tip of the string, i.e. the point where
the closed string folds on itself,
falls in the radial direction of AdS on a null geodesic.

Being in the tensionless limit, we can make the total
energy of the string arbitrarily small and hence
the geometry is not back reacted. Time evolution will
solely be the result of causal effects.
The light cone of the tip of the falling string
will divide the bulk points into those
that "know" of the formation of a closed string and those that do not. The light cone
is the holographic extension of the {\it{quasi-particles}}
that are produced in the process of quench
and which propagate in both directions on the line and change the
state of the field theory.

Entanglement Entropy of various subspaces are described holographically
by the Ryu-Takayanagi (RT) curves \cite{Ryu:2006bv}-\cite{Ryu:2006ef}.
We propose how the light cone of the falling string deforms
these curves and hence find the time dependent
EE in terms of lengths of various curves in the bulk.
Our results produce the field theory expressions
in full detail.

The rest of the paper is organized as follows. In the next section we briefly
review various ingredients that we need for the rest of the work.
In particular we give a sketch of the field theory treatment of quench,
mention the previous attempts for a holographic description of this process,
state the AdS/BCFT correspondence and review the Yo-Yo string solution.
In section three we state our proposal for holographic quench. In section
four we calculate EE following a local quench by using the deformed RT curves.
We conclude with some remarks
and discussions and work out some details in the appendix.
\\
%%%%%%%%%%%%%%%%%%%%%%%%%%%%%%%%%%%%%%%%%%%%%%%%%%%%%%%%%%%%%%%%%%%%%%
%%%%%%%%%%%%%%%%%%%%%%%%%%%%%%%%%%%%%%%%%%%%%%%%%%%%%%%%%%%%%%%%%%%%%%
\section{Review Material}
In this section we review different topics that we will need for our treatment
of local quench by holography. We will be very brief in each part and simply state
the main results.
\\
%%%%%%%%%%%%%%%%%%%%%%%%%%%%%%%%%%%%%%%%%%%%%%%%%%%%%%%%%%%%%%%%%%%%%
\subsection{Local Quench by Field Theory}\label{FT}
The main reference for this work is \cite{Calabrese:2006rx}.
The approach here is to find
the path integral representation for the density
matrix of the system at a time $t$ after
quench. This amounts to considering the two world
sheets with boundaries which have
evolved from $t=-\infty$ to $t=0$. The two world sheets
are then joined together, to produce one
with no boundaries now, and is allowed to evolve
for another time period $t$.
The path integral on the resulting manifold
corresponds to the "ket" of the
system at time $t$ after quench.

A similar manifold will then correspond to the
"bra" state and, once put together,
the path integral on the two manifolds glued
together will represent the density matrix.
One can then calculate the expectation value o
f various operators by introducing
local operator insertions on the manifold. To make the
expressions convergent a regulator
, $\epsilon$, is introduced in the path integral.
Once analytically continued, the
whole thing will correspond to a path integral on a
plane with two cuts along imaginary time,
one from $-\infty$ to $-i\epsilon$ and the other from
$i\epsilon$ to $+\infty$.

The final step is to use a conformal transformation to map
the manifold with two cuts, parameterized
by $(z,\bar{z})$, into the upper half plane of
$(w,\bar{w})$ with $Re(w)>0$. This is achieved by

\begin{equation}\label{wzmap}
z\xrightarrow{f}w(z)=\frac{z}{\epsilon}+\sqrt{1+\left(\frac{z}{\epsilon}\right)^2}
\,\,\,\,\,\, \text{with inverse}\,\,\,\,\,\, w\xrightarrow{f^{-1}}z(w)=\epsilon\frac{w^2-1}{2w}\, .
\end{equation}

To calculate, for example, time dependent entanglement entropy of
various intervals, one should insert the so called twist operators
at proper locations and calculate the corresponding
$n$-point functions. This amounts to doing the standard
calculations on the upper half plane and use
the conformal map to find the expressions on the original manifold with cuts.

This procedure has been done with great care and detail in \cite{Calabrese:2006rx}.
In particular the time dependent EE following a local quench has
been calculated for an interval with end points at different
locations.
%%%%%%%%%%%%%%%%%%%%%%%%%%%%%%%%%%%%%%%%%%%%%%%%%%%%%%%%%%%%%%%%%%%%%%%%%%%%
\subsection{AdS/BCFT Correspondence}\label{ADSBCFT}
The main references here are \cite{Takayanagi:2011zk},\cite{Fujita:2011fp}.
Consider a CFT living on a manifold $\M$ with boundary $\partial\M$. The
conjectured AdS/BCFT correspondence states that this system has a
gravitational dual consisting of a part of AdS geometry $\N$,
together with a co-dimension one hypersurface $Q$, such that
$\p\N=\M\cup Q$. It is important to note that $Q$, which is
homologous to $\M$, is part of the holographic description.

\begin{figure}[H]
\begin{center}
\includegraphics[scale=0.2]{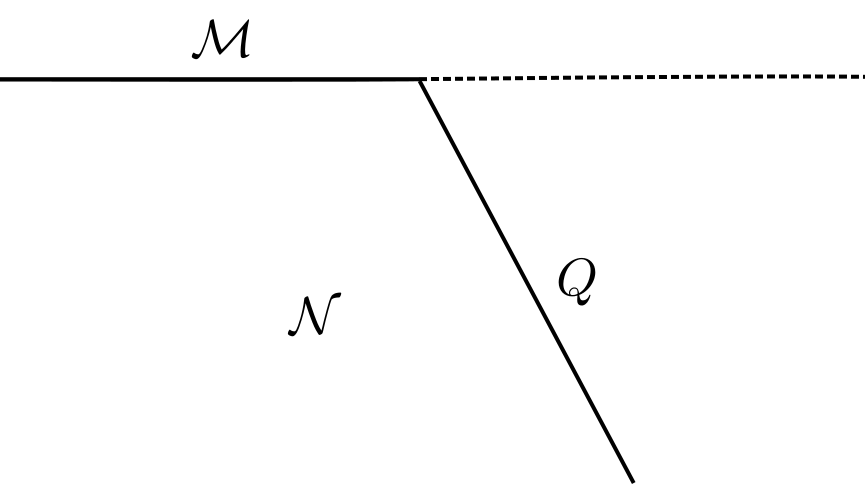}
\end{center}
\caption{Holographic picture of a BCFT defined on the half-plane, $\M$.}
\label{figA1}
\end{figure}

The action that describes this system is
\begin{equation}
S=S_\N +S_Q\ ,
\end{equation}
where
\begin{equation}\label{SQ}
S_\N =\frac{1}{16\pi G_N}\int_\N d^d x \sqrt{-G}(R-2\Lambda)\ ,\,\,\,\,
S_Q =\frac{1}{8\pi G_N} \int_Q d^{d-1}x\sqrt{-h}(\K-T)\ .
\end{equation}
In these expressions $h_{ij}$ and $\K_{ij}$ are the induced metric and
extrinsic curvature of
$Q$ respectively and $\K=h^{ij}\K_{ij}$.
The constant $T$ is the tension of $Q$. One should also include
 the boundary action associated with $\M$
\begin{equation}
 S_\M=\frac{1}{8\pi G_N} \int_\M d^{d-1}x\sqrt{-g}K\ ,
\end{equation}
where again $g_{ij}$ and $K_{ij}$ are the induced metric and
extrinsic curvature of $\M$, respectively
and $K=g^{ij}K_{ij}$.
We choose the usual Dirichlet boundary condition on
$\M$ but Neumann boundary condition on $Q$.
Then variational principle gives us Einstein equation
in the bulk region, $\N$,
as well as the following constraint on the boundary $Q$
\begin{equation}
\K_{ij}=h_{ij}(\K-T)\ .\label{braneeqm}
\end{equation}
The case of our interest, which we will need in future sections,
is the dual description of a $1+1$ dimensional
BCFT on the half plane $x<0$. Using the Poincar\'e
coordinates for AdS$_3$ as
\begin{equation}\label{Poincare metric}
ds^2=\frac{L^2}{z^2}\left(-dt^2+dz^2+dx^2\right)\ ,
\end{equation}
we can parameterize The hypersurface $Q$ as $x_Q=x_Q(z)$.
The unit normal on this surface reads
\begin{equation}
n^\mu =\left(n^t,n^z,n^x\right)=\frac{z}{L\sqrt{1+{x'_Q}^2(z)}}
\left(0,-x'_Q(z),1\right)\, .
\end{equation}
Assuming a constant $T$ for simplicity, equation \eqref{braneeqm}
determines the profile of $Q$
\begin{equation}\label{Qsurface}
x_Q(z)=\alpha(T)\, z+\beta\, ,
\end{equation}
where
\begin{equation}\label{Qsurface1}
\alpha(T)\equiv \frac{TL}{\sqrt{1-T^2L^2}}\, ,
\end{equation}
and since $x=0$ when $z=0$, we should set $\beta=0$.
The situation has been schematically shown in figure (\ref{figA1})

One last point is that the boundary entropy of the BCFT is related
to $T$. In particular, when $T=0$, the boundary entropy also vanishes.
%%%%%%%%%%%%%%%%%%%%%%%%%%%%%%%%%%%%%%%%%%%%%%%%%%%%%%%%%%%%%%%%%%%%%%%%%%%%%%%%
\subsection{The Yo-Yo String}\label{yoyo}
The main reference here are \cite{Bardeen:1975gx} and \cite{Ficnar:2013wba}. The Yo-Yo solution is
an example of a classical motion of an open string
with finite momenta at end points. The main feature of such
strings is that the end points move on light-like
geodesics on a direction along the string itself.
This is not compatible with any of the two allowed boundary conditions for open strings,
namely the Dirichlet and Neumann boundary conditions. Still, they can be found as certain limits
of allowed string configurations and hence should be considered as part of the classical theory.

In flat space, one can write solutions that interpolate between the Yo-Yo and the Regge solutions, the latter
being an open string which rotates as a rod with a constant angular velocity. Such solutions
satisfy the Neumann boundary conditions at ends and describes open strings that while rotating,
their size changes between two positive numbers $l_1\ge l_2>0$. At any instant of time, the string is a straight line.

The Regge limit is when $l_1=l_2$. The other limit, Yo-Yo limit, is when $l_2\rightarrow0$. In the Yo-Yo limit the 
Neumann boundary condition breaks down\footnote{It can be easily seen in the static gauge.} and for the energy of the string to be conserved one has to add
a boundary term to the usual string action. The whole process can be generalised to arbitrary curved backgrounds
\cite{Ficnar:2013wba}.

In theories where open strings are not allowed, one can produce closed strings with the same above features.
This amounts to attaching two open strings at end points and produce a closed string, folded back on itself, which, at any instant of time
looks like a straight line. The important difference is that the open string ends are now replaced by the folding points 
which do not have to satisfy the open string boundary conditions. 
In the Yo-Yo limit, we will no longer need the additional boundary term and each half of the closed string provides the other half
with the required boundary conditions\footnote{In fact the two boundary terms for each half appear with opposite signs due to
the orientation of the boundary and cancel out}. Therefore one has valid classical solutions of strings which carry momentum at 
the folding points. In the AdS background, these configurations are special cases of the rotating folded strings which 
were studied in \cite{Gubser:2002tv}, when the angular momentum is zero. The configurations are uniquely specified by a single number, the total length of the string or 
equivalently the total energy.

The solution of our interest in this work is the folded closed string in AdS$_3$
in coordinates (\ref{Poincare metric}).
Working out the details (look at \cite{Ficnar:2013wba})
one finds that the folding points move on the trajectory $z=t$.
Choosing the direction $z$ to parameterize
this trajectory, one finds that the momentum of the folding point evolves as
\begin{equation}\label{yoyoevolution}
\dot{p}_t=-\frac{TL^2}{4\pi G_N}\frac{1}{z^2}\ ,
\end{equation}
where dot stands for differentiation with respect to $z$.
One can state the initial value of $p_t$ in terms of the total
energy of the string, $E$, which is $p_t(0)=-E/2$. Therefore one finds
\begin{equation}\label{yoyomomentum}
p_t=-\frac{E}{2}+\frac{TL^2}{4\pi G_N}\frac{1}{z}\ .
\end{equation}
Using this relation one can find the point where the folding point changes direction
\begin{equation}\label{snap}
z_*=\frac{TL^2}{2\pi G_N E}\ .
\end{equation}
This is called the ``snapback" point and is the closest the folding point can
get to the boundary at $z=0$.
This relation will find special significance in next section when we
propose our holographic model for local quench.
%%%%%%%%%%%%%%%%%%%%%%%%%%%%%%%%%%%%%%%%%%%%%%%%%%%%%%%%%%%%%%%%%%%%%%%%%%%%%%%%%%%%%
\subsection{Quench By Holography}
There have been a number of attempts to describe local quench by holography,
\cite{Nozaki:2013wia}, \cite{Ugajin:2013xxa} and \cite{Asplund:2013zba}. see also \cite{Basu:2011ft}-\cite{Roberts:2012aq}. Here we mention two of these works which are of more relevance to ours.

The first one is \cite{Nozaki:2013wia}. The process that is studied in this work is
when a CFT undergoes a localized excitation. The excitation spreads throughout the
system and changes the original state. The holographic model that \cite{Nozaki:2013wia}
proposes for this process is to consider a massive particle which is falling freely
in the AdS spacetime. As a result, the geometry in the bulk will be affected and
the state of the system changes in time. Of particular interest is to compute
the time dependence of EE for various subspaces. This is achieved by applying
the RT prescription to the backreacted geometry and finding the corresponding
minimal hypersurfaces in the bulk. In the case of AdS$_3$ the
calculations can be done analytically. The results are in general
agreement with expectations from
field theory calculations.

The second work we mention here is \cite{Ugajin:2013xxa}, see also \cite{Asplund:2013zba}.
This reference makes a parallel computation of section(\ref{FT})
in the bulk. In particular, the bulk version of the map (\ref{wzmap})
is introduced and the field theory correlation functions of the
twist operator are calculated by the usual RT curves in the bulk.
The results thus obtained reproduce the field theory calculations of
\cite{Calabrese:2006rx} correctly.

In the next section we propose another model for local quench when
two CFT's on semi-infinite lines are joined together.
Our model considers each of the original BCFT's in terms of their dual
descriptions through the AdS/BCFT correspondence. Local quench occurs when the two
extensions of the field theory boundaries, i.e., the $Q$ surfaces mentioned in (\ref{ADSBCFT})
detach from the boundary and attach to one another. Their subsequent evolution in bulk determines
the time dependence of the process. In this work we only consider BCFT's with a zero
boundary entropy. Extension to a general case is postponed to future works.
%%%%%%%%%%%%%%%%%%%%%%%%%%%%%%%%%%%%%%%%%%%%%%%%%%%%%%%%%%%%%%%%%%%%%%%%%%%%%%%%%%%%%%%
%%%%%%%%%%%%%%%%%%%%%%%%%%%%%%%%%%%%%%%%%%%%%%%%%%%%%%%%%%%%%%%%%%%%%%%%%%%%%%%%%%%%%%%
\section{Local Quench By AdS/BCFT and Yo-Yo String}\label{model}
Consider a CFT living on a semi-infinite line $\M$ with $x<0$. The dual description will be
the one depicted in figure (\ref{figA1}). The $Q$ surface is determined by equations
(\ref{Qsurface}) and (\ref{Qsurface1}). Specify to the situation when the tension on $Q$
is zero. In the field theory language this means that the boundary entropy for this system is zero.

A key observation is that for $T=0$, the surface $Q$ is simply the world sheet of
an open string which satisfies the classical equations of motion for a bosonic
Polyakov string. The string attaches the boundary at $z=0$ with the usual Neumann
boundary condition. The string, however, is tensionless; $\alpha'\rightarrow\infty$. To see this,
one can simply note that for $T=0$, the extrinsic curvature $\K$ of the surface $Q$ vanishes
and hence it is a minimal surface. Classical solutions for Polyakov action are also
nothing but minimal world sheets. The Neumann boundary condition is also
obvious from (\ref{Qsurface}) which gives
\begin{equation}
\partial_zx_Q|_{z=0}=0\ \ \ \ \  \text{when}\ \ \ T=0\ .
\end{equation}

\begin{figure}[H]
\begin{center}
\includegraphics[scale=0.4]{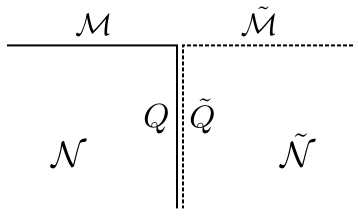}
\end{center}
\caption{Two disconnected BCFT's when $T\rightarrow 0$}
\label{fig1}
\end{figure}

Now consider an identical field theory which lives on the semi-infinite line
$\tilde{\M}$ with $x>0$. Again the same situation holds as above.
Figure (\ref{fig1}) shows the two systems together with $Q$ and $\tilde{Q}$
being the world sheets of tensionless Polyakov open strings

Local quench occurs as a result of manipulations at and around the point $x=0$ which
joins the two disconnected field theories into a single one. The question is what
should be considered as the bulk version of these manipulations.  Whatever these 
may be, they should occur in the UV part of the AdS space, around the
$x=0$ point, which is a result of (rather) local manipulations around this point. It should
ultimately result in the removal of the partition between the initially separate systems.
Mirroring the field theory process, this should happen in a time dependent manner which 
entangles the degrees of freedom of the two systems as time goes by until one ends up 
in the ground state of single system, consisting of the two. Since this state is the ground state
of conformal field theory on a line, the dual geometry must be empty AdS space.

\begin{figure}[H]
\begin{center}
\includegraphics[scale=0.4]{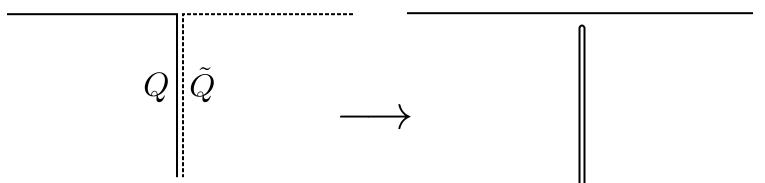}
\end{center}
\caption{Holographic picture for local quench; appearance of a folded string}
\label{fig2}
\end{figure}

Our proposal is that
the bulk version corresponds to detaching each of the open strings from the boundary and
attaching the two ends together. This will produce a folded closed string
that can now propagate in the bulk  (see figure(\ref{fig2}))\footnote{The other end of the string is behind the
Poincar\'e horizon and we do not worry about it. If we consider global AdS, a similar
process happens in the other end and two BCFT's on intervals join into one on a circle.
In bulk, a closed string forms which becomes point-like in late times. If one insists
to have BCFT's on semi-infinite lines, one can still work in global coordinates but remove
one point from the angular direction. The picture will then describe an open string
which is folded in bulk with end points attached to boundary at infinity with Neumann
boundary condition. This will not affect the motion of the
folding point.}. This propagation happens according to the classical equations of motion
for the string and should eventually result in the removal of the string, or its shrinking to a point,
such that we are left with an empty AdS space at late times.

To find the subsequent motion of the string we note that once we attach the endpoints, we are left with a folded closed string
which stretches along the radial coordinate in AdS with all of its points, including the folding point, at rest with zero momentum.
This coincides with a tensionless yo-yo string at the snapback point which, in turn,  is uniquely specified by its initial conditions. The dynamics will
thus be governed by equations (\ref{yoyoevolution}) and (\ref{yoyomomentum}).
The only remaining initial condition is the total energy of the string or the $z$ value of the folding point.

To find this last piece of information to determine the motion of the string we should properly 
define the tensionless limit of our yo-yo. Consulting equation  (\ref{snap}), we see that a finite value for the snapback point
$z_*$ can be find by
\begin{equation}\label{yoyolimit}
E\rightarrow 0\ ,\,\,\,\,\, T\rightarrow 0\ ,\,\,\,\,\,
\lim_{T\rightarrow 0, E\rightarrow 0}z_*=\lim_{T\rightarrow 0,
E\rightarrow 0}\frac{TL^2}{2\pi G_N E}\equiv\delta\ .
\end{equation}
This is a reasonable limit once we remember what case of a quench we are studying. We are attaching two BCFT's 
with {\it{zero boundary entropy}}. In such a case no energy is released or absorbed by quench \cite{Calabrese:2006rx}.
Holographically this means that our AdS space never suffers back reactions by the folded string. This is also reflected
in the $Q$ surfaces which are tensionless. 

The last ingredient in our holographic picture for quench is to relate $\delta$ defined in (\ref{yoyolimit}) to
field theory. 
A relevant question in the field theory side is how local the manipulations have been.
In other words to what extent have the adjacent points around $x=0$ have been perturbed
or excited as a result of quench. Let us assume that the process of attaching the two field theories has
affected a region of size $\epsilon$.

\begin{figure}[H]
\begin{center}
\includegraphics[scale=0.4]{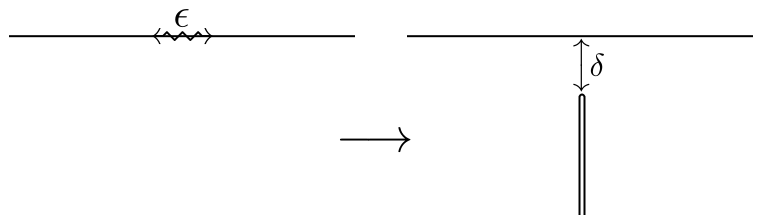}
\end{center}
\caption{Regulators in boundary (Left) and bulk (Right).}
\label{fig3}
\end{figure}

We suggest that an interpretation of this length in terms of the bulk physics comes through
the snapback point $z_*$ and hence $\delta$ (see figure (\ref{fig3}))

\begin{figure}[H]
\begin{center}
\includegraphics[scale=0.4]{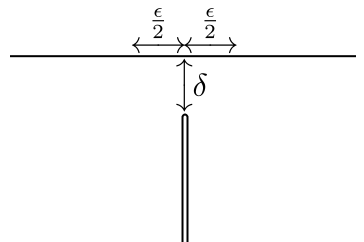}
\end{center}
\caption{Identification of the regulators.}
\label{fig4}
\end{figure}

A natural identification for our holographic picture will be then (fig.\ref{fig4})

\begin{equation}\label{epsilondelta}
\delta=\frac{1}{2}\ \epsilon\ .
\end{equation}

Let us see how this picture produces the general features of quench in the bulk.
Dealing with zero boundary entropy for the initial BCFT's, the whole process
is governed by causal effects rather than energy transfer. In the field theory
language this is a result of propagation of {\it{quasi-particles}}
in both direction on the line which deliver the message of manipulations at $x=0$.
These quasi-particles have an arbitrarily small energy.

\begin{figure}[H]
\begin{center}
\includegraphics[scale=0.3]{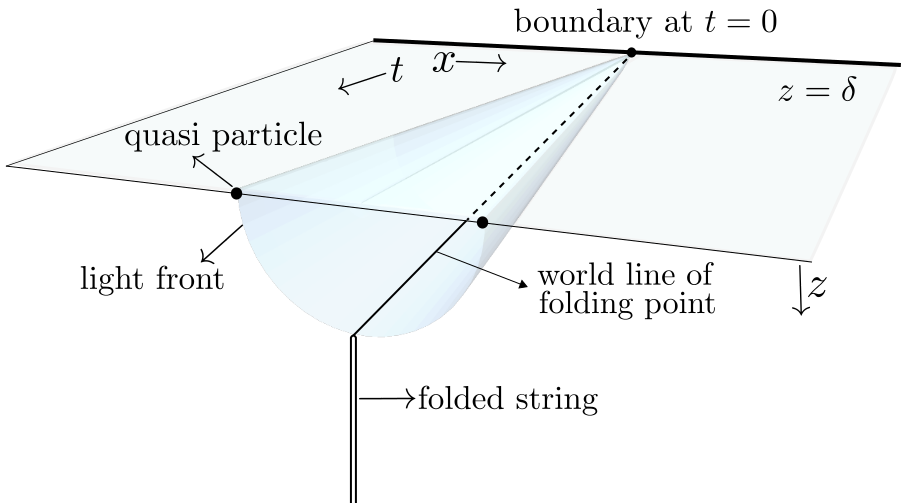}
\end{center}
\caption{Holographic local quench: Light-cone and $\delta$ Vs. quasi-particles and $\epsilon$.}
\label{fig5}
\end{figure}

In the bulk, the folded string falls freely away from the boundary into the bulk.
The folding point falls on a light-like geodesic. The string has an arbitrarily small
energy and does not back react on the geometry, again all that comes into play
is the causal effects. The message of formation of the folded string propagates into
the bulk along the light-cone depicted in figure (\ref{fig5}).

The light-cone divides bulk points into those who {\it{know}} of the manipulations
and those who do not. The light front is identified as the bulk version of
quasi-particles (\ref{fig5}).

In the next section we will put our model to test by studying the time evolution
of entanglement entropy of a single interval after quench.

%%%%%%%%%%%%%%%%%%%%%%%%%%%%%%%%%%%%%%%%%%%%%%%%%%%%%%%%%%%%%%%%%%%%%%%%%%%%
%%%%%%%%%%%%%%%%%%%%%%%%%%%%%%%%%%%%%%%%%%%%%%%%%%%%%%%%%%%%%%%%%%%%%%%%%%%%

\section{Entanglement Entropy and Local Quench}
The time dependence of entanglement entropy after quench
will provide us with information of how the initial ground states of the
disjoint BCFT's will evolve into that of a single CFT on a line.
Let us denote the entangling subregion by $A$ and its entanglement entropy by $S_A$.
The complementary region and its entropy are denoted by
$\bar{A}$ and $S_{\bar{A}}$ respectively.
Since the entire system is in a pure state at any instant of
time, $S_A=S_{\bar{A}}$ remains true at all times.

In field theory, one starts with the initial value for $S_A$.
As the quasi-particles penetrate $A$, the entanglement
pattern begins to change. One again arrives at a final
value for $S_A$ as the quasi-particles exit $A$.

In the bulk side, one again starts with an initial condition. Denoting
the endpoints of $A$ by $(x_i,x_j)$, the holographic
value for $S_A$ is given by the RT curve through
\begin{equation}\label{RT prescrip}
S_{A}=\frac{\CA(\gamma_{x_ix_j})}{4G_N}\, ,
\end{equation}
where $G_N$ is the newton constant, $\gamma_{x_ix_j}$ is
the geodesic curve in the bulk that is homologous to $A$ and
$\CA(\gamma_{x_ix_j})$ is the length of $\gamma_{x_ix_j}$.
As the light front of figure (\ref{fig5}) intersects with $\gamma_{x_ix_j}$,
it deforms the standard RT curve and changes $S_A$.
Below, we propose how these deformations and modifications take place
and try to justify them by physical intuitions and arguments.

The example we pick up for illustration is when $A$ is
entirely inside one of the original
BCFT's. We follow the conventions of \cite{Calabrese:2006rx}
to make comparisons immediate. The endpoints of $A$ are
located at $\ell_2$ and $\ell_1$ with $\ell_1>\ell_2>0$.
The two BCFT's are joined at $x=t=0$. We first recall
how the RT prescription works before quench and then
propose how to modify it.

\subsection{Entanglement entropy before the quench}

%%%%%%%%%%%%%%%%%%%%%%%%%%%%%%%%%%%%%%%%%
To apply the standard RT prescription for holographic Entanglement Entropy 
one should find among all the curves that are homologous to the entangling region 
in the boundary the one that has the minimum length. In the case of a BCFT
where additional boundaries exist both in the bulk and field theory sides the prescription
has to be modified.

This amounts to considering two different types of curves, connected and disconnected.
The former are the usual RT curves that end on the boundaries of the entangling region
while the latter can end on the Q-surface in the bulk. Depending on the configuration of 
the entangling region and the Q-surface, the connected or disconnected pieces yield the 
minimum length and hence the EE.

To be more specific we consider the setup in the previous section where we have 
two BCFT's with a common boundary and with a zero boundary entropy each. 
In the symmetrical situation at hand we find the introduction of "\emph{Image Points}" of
special interest and use.\footnote{In non symmetrical situations the
image points can also be used.} These points make the application of the RT prescription
more easy and systematic. We will later find that any modifications to the standard
recipe will be facilitated by the image points.

One can see that the connected RT curves, $\gamma_c$, are those that start and end on (image) points 
whereas the disconnected curves, $\gamma_d$, correspond to those that have a point and an image
on the ends. Below in figure \eqref{fig15}, we have depicted the connected vs. disconnected RT curves.

\begin{figure}[H]
\begin{center}
\includegraphics[scale=0.35]{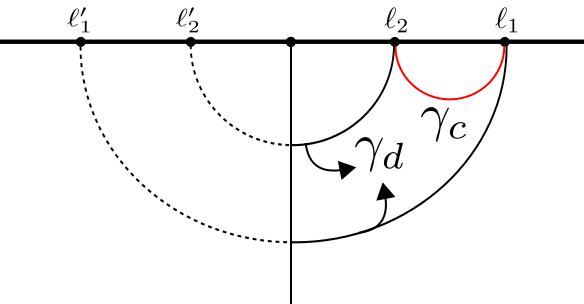}
\end{center}
\caption{Connected vs. disconnected geodesic curves.}
\label{fig15}
\end{figure}

 The recipe can then be summarized as\footnote{We will drop the factor $1/4G_N$ in Eq. \eqref{RT prescrip}, henceforth. }
\be\label{cd proposal}
S_{EE}=\text{Min}\lbrace\A(\gamma_c)\, , \, \A(\gamma_d) \rbrace\, ,
\ee

There are a number of comments worth mentioning. The problem of a single interval
as the entangling region in a BCFT is in effect that of two disjoint intervals in a normal
CFT. This is well expected as in a BCFT, the calculation of the two
point function of twist operators will in effect be that of a four point function 
of operators inserted at points and their images, see for example \cite{DiFrancesco:1997nk}. In the holographic setup this
corresponds to the curves that can end on image points. 

It is well known that four point functions cannot be completely fixed by conformal symmetry
completely. They depend also on the operator content of the theory. 
They, however, include a universal part which is fixed by symmetries. Using the notation of \eqref{wzmap} we can find the correlation function of twist operators, $\sigma$ and $\tilde\sigma$, and thus the partition function of the replicated theory as follows
\be
Z_n=\Tr \rho_A^n\propto \langle\sigma\tilde{\sigma}\text{(interval endpoints)}\rangle^n_z\sim \langle\sigma\tilde{\sigma}\text{(interval endpoints)}\sigma\tilde{\sigma}\text{(image points)}\rangle^n_w\, .
\ee
Consequently, we will have
\be
Z_n= Z_n^{(u)}\, \widetilde{\F}_{n}(\{x\})\, ,
\ee
where $Z_n^{(u)}$ represents the universal part of the replicated partition function, and $ \widetilde{\F}_{n}(\{x\})$ stands for a non-universal function of cross-ratios which are shown by $\{x\}$ in sum.

However, as we will argue later, the important part for our purposes is the universal one. Using this technique the authors of \cite{Calabrese:2006rx} have computed the EE for various cases. For the most general case, i.e. for a single interval $A:[\ell_2>0,\ell_1>0]$ they have found
\begin{equation}
\label{EEformula}
S_{A}=\frac{c}{6}\log \frac{(\ell_1-\ell_2)^2}
{(\ell_1+\ell_2)^2}\frac{4\ell_1\ell_2}{a^2}\, .
\end{equation}

To facilitate the following discussion, and also to make contact with the holographic description of EE, we rewrite the above relation as a sum of separate parts
\begin{equation}
\label{EEseparate}
S_{A}=S_c+S_d+S_n=\frac{c}{6}\log \frac{(\ell_1-\ell_2)^2}{a^2}+\frac{c}{6}\log \frac{4\ell_1\ell_2}{a^2}+\frac{c}{6}\log \frac{a^2}{(\ell_1+\ell_2)^2}\  ,
\end{equation}
where the subscripts $c, d, n$ refer to connected, disconnected and negative contributions respectively.
We recognize the first two contributions as corresponding to the familiar connected, $\gamma_c$, and 
disconnected, $\gamma_d$, RT curves. The negative contribution, however, although appearing 
in the universal part, has no counterpart amongst the RT curves.

In order to get an insight into this contribution to EE, consider a situation where one 
changes the length of the entangling region or its position with respect to the boundary 
smoothly. Interestingly, the negative contribution can make the transition between 
connected and disconnected pieces a smooth one. To see this we first consider
the limit $\ell_1\gg \ell_2>0$ in (\ref{EEformula}). In this limit $\ell_1+\ell_2\sim \ell_1-\ell_2$ and 
the disconnected piece, $\gamma_d$,  dominates. In the opposite limit $\ell_2\gg \ell_1-\ell_2>0$,
we have $\ell_1\sim\ell_2$ and thus $(\ell_1+\ell_2)^2\sim4\ell_1\ell_2$ and therefore
$\gamma_c$ survives.

Moving to the holographic description we find it natural to include the negative contribution in terms of geodesic
curves specially since it appears in the universal part. This should be understood as our proposal for how to 
modify the RT prescription in case of multiple intervals as well as a single interval in presence of a boundary.

The holographic counterparts of these contributions can easily be identified as 
two intersecting curves between points and their images. Denoting the image points by $\tilde{\ell_1}$ and  $\tilde{\ell_2}$ the negative contribution 
can be produced by
\be
\text{Negative contribution}:\frac{1}{2}\{\CA(\gamma_{\ell_1\ell'_2})+\CA(\gamma_{\ell'_1\ell_2})\}\, .
\ee 
 So in summary, we have proposed the following prescription for computing the entanglement entropy holographically, before quench
\begin{equation}
\label{stRT}
\CA_0=\frac{1}{2} \{\CA(\gamma_{\ell_1\ell_2})+\CA(\gamma_{\ell'_1\ell'_2})\}+\frac{1}{2}
\CA(\gamma_{\ell_1\ell'_1})+\frac{1}{2}\CA(\gamma_{\ell_2\ell'_2})-\frac{1}{2}
\{\CA(\gamma_{\ell_1\ell'_2})+\CA(\gamma_{\ell'_1\ell_2})\}\ .
\end{equation}
This coincides with the universal part of the field theory calculations, produces the
 connected and disconnected pieces in the appropriate limits and makes a transition
 between the limits possible. (\ref{stRT}) should be considered as our proposal to correct/modify the standard 
 recipe of (\ref{cd proposal}).

\begin{figure}[H]
\begin{center}
\includegraphics[scale=0.3]{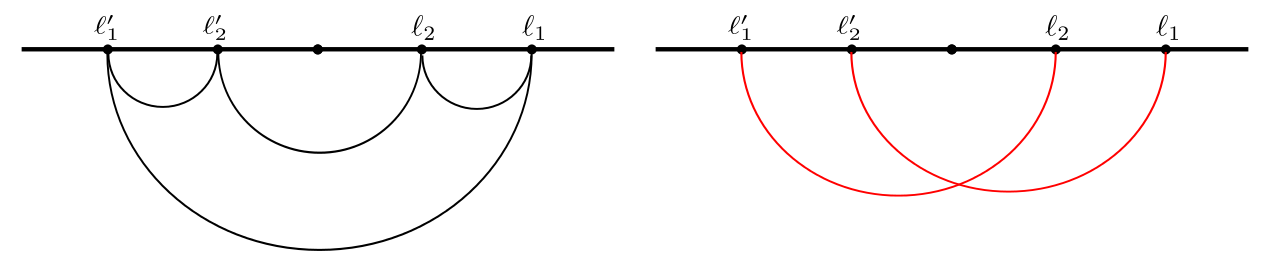}
\end{center}
\caption{Positive (Left) vs. Negative (Right) contributions.}
\label{fig6}
\end{figure}

We have intentionally separated the contributions from points and their
images because, as we will see in the following, the modifications
will treat them differently. In figure (\ref{fig6}) we have shown
the positive and negative contributions by black and red
curves, respectively.

In sum we believe that \eqref{stRT} should be considered as the holographic prescription for computing the EE which sums two connected and disconnected pieces and includes the negative contribution. We would like to emphasize again that it is somehow different from \eqref{cd proposal}, since in the latter, among the connected and disconnected pieces only the minimal one is taken into account. 

Incidentally we recently became aware of  \cite{Hubeny:2007re} in which the authors have computed the EE for two disconnected subregions in $1+1$ dimensions. Interestingly, our result is in a complete agreement with what they have found and hence this would provide a further support for what we have proposed in eq.(\ref{stRT}).\footnote{We would like to thank the referee for bringing this reference to our attention.}

The above prescription will give us the following value for $S_A$
\begin{equation}\label{stRT1}
S_{A}(t<\ell_2)=\frac{c}{6}\log \frac{(\ell_1-\ell_2)^2}
{(\ell_1+\ell_2)^2}\frac{4\ell_1\ell_2}{a^2}\, ,
\end{equation}
where $c$ is the central charge of the CFT and $a$ is a
regularization parameter which is the lower
bound for the coordinate $z$ (look at Appendix (\ref{appA}) for details).

\subsection{Entanglement entropy after the quench}
As stated before the main reference for field theory results is \cite{cc0711}. As we discussed before, in their
calculations universal as well as non-universal parts appear. The latter are in general,
functions of cross ratios. In some limits these contributions become negligible.

As argued by the authors of \cite{cc0711}, the EE after quench is well approximated by the 
universal parts. The non-universal parts are suppressed except at special times when
the quasi-particles pass through the boundaries of the entangling region, i.e.,
the end points of the intervals.\footnote{We would like to thank John Cardy for his useful comment on this point.}

In the following where we propose how the RT curves must be deformed and modified 
after quench, we are able to produce the universal contributions to EE as a function
of time. This means that at times when the non-universal parts become important, 
further modifications should be introduced. Whatever these might be, they should
depend on the specific CFT under study.

In the limit of a large central charge where classical gravity is believed to 
capture universal features of CFT's, one may naturally expect that non-universal
contributions will not have a simple gravitational description.

First step is to motivate the modification we formerly pointed out and to do that we note that the complementary region to $A$,
denoted by $\bar{A}$, changes in time. At early times when $A$ has not yet
received the message of quench, its reduced density matrix $\rho_A$ is
unaffected and hence $S_A$ remains unchanged. After the quasi-particle
has penetrated $A$ (say at time $t=t_*$), part of the region, denoted by $ \alpha\subset A$,
receives the message and the entanglement pattern begins to change. This continues
until the quasi-particle exits the interval when $S_A$ assumes its steady state value.
We are only interested in the intermediate times $\ell_2<t<\ell_1$ in the following.

Entanglement entropy changes by two competing contributions; one that
decreases $S_A$ and one that increases it.
Some of the existing entanglements
disappear and some new ones form.
In other words, the entanglement between $A$ and $\bar{A}(t<t_*)$ transfers
to that between $A$ and $\bar{A}(t>t_*)$.

As the quasi-particle pair travel in both directions,
the degrees of freedom in $\alpha$ find new counterparts to entangle with. These are
those degrees of freedom which have been swept by the pair. This in turn transfers
part of entanglement between $A$ and $\bar{A}(t<t_*)$ to $\alpha$ and $\bar{A}(t>t_*)$.

\begin{figure}[H]
\begin{center}
\includegraphics[scale=0.3]{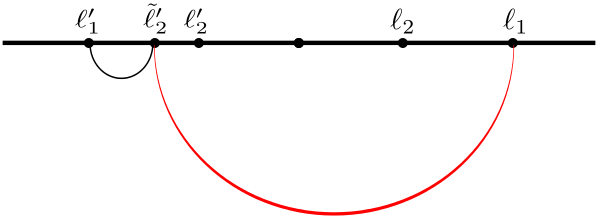}
\end{center}
\caption{Image point moved to left. This decreases entanglement.}
\label{fig7}
\end{figure}

The decreasing contribution can be best understood by focusing on the image point $\ell'_2$.
Note that quasi-particles deal with points and their images differently. This is because the image points
owe their existence to the boundary and as the boundary is removed the image points and their effects
should be swept away. It is thus reasonable to assume that the point $\ell'_2$
is swept off to left to $\tilde{\ell}'_2$ (see figure (\ref{fig7})) as the quasi-particle passes through it.

Turning to the holographic description, this will modify some of the original RT curves. Out of the six contributions in (\ref{stRT}), let us first focus on the four curves
$\gamma_{\ell_1\ell_2}, \gamma_{\ell'_1\ell'_2}, \gamma _{\ell_1\ell'_2}$ and $\gamma_{\ell_2\ell'_1}$ and leave the remaining two for later.
The shift of $\ell'_2\rightarrow\tilde{\ell}'_2$ decreases the positive
contribution of $\CA(\gamma_{\ell'_1\ell'_2})\rightarrow \CA(\gamma_{\ell'_1\tilde{\ell}'_2})$
and increases the negative contribution of $\CA(\gamma_{\ell_1\ell'_2})\rightarrow \CA(\gamma_{\ell_1\tilde{\ell}'_2})$
while leaves the two parts $\CA(\gamma_{\ell_1\ell_2})$ and $\CA(\gamma_{\ell_2\ell'_1})$ unchanged.

The modified curves, up to now,  are summarized in figure(\ref{fig7}).

\begin{figure}[H]
\begin{center}
\includegraphics[scale=0.3]{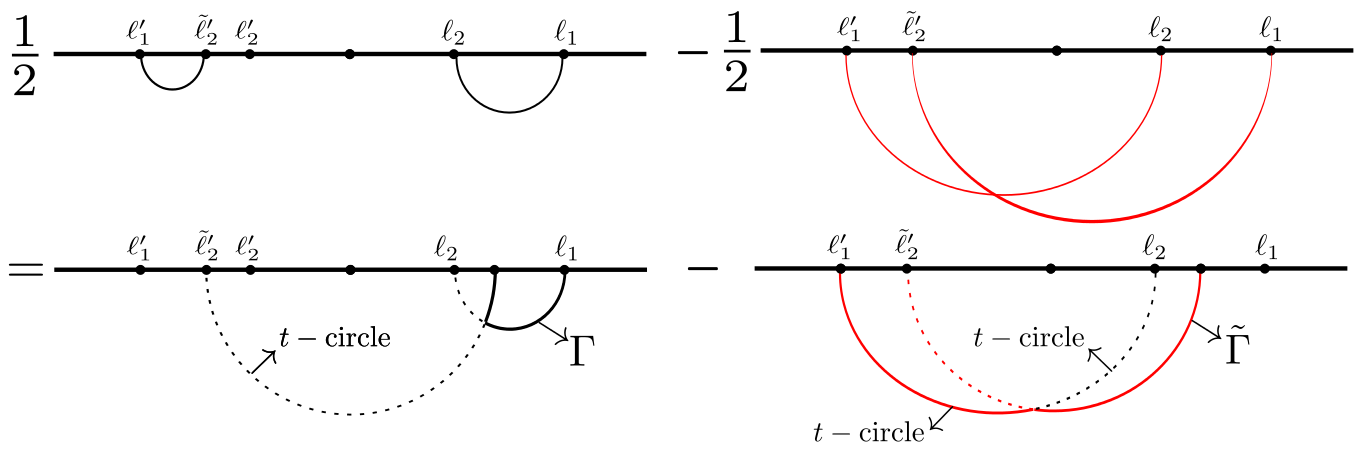}
\end{center}
\caption{RT curves deformed by light-cone: $\Gamma$ and $\tilde{\Gamma}$.}
\label{fig8}
\end{figure}

Using RT formula for the deformed curves and noting that $\tilde{\ell}'_2=t$, we have
\begin{equation}
\label{decrease1}
\CA(\gamma_{\ell'_1\tilde{\ell}'_2})=\frac{c}{6}\log\frac{(\ell_1-t)^2}{a^2}\ ,\,\,\,\,\,
\CA(\gamma_{\ell_1\tilde{\ell}'_2})=\frac{c}{6}\log\frac{(\ell_1+t)^2}{a^2}\ .
\end{equation}
The contribution of the mentioned four curves appears in the following combination
\begin{equation}
\label{decrease2}
\frac{1}{2}\{\CA(\gamma_{\ell_1\ell_2})+\CA(\gamma_{\ell'_1\tilde{\ell}'_2})\}-
\frac{1}{2}\{\CA(\gamma_{\ell_1\tilde{\ell}'_2})+\CA(\gamma_{\ell'_1\ell_2})\}\ ,
\end{equation}
Quite interestingly, this combination can be depicted in terms of geodesic curves in a very suggestive way.
This is shown in figure (\ref{fig8}) and illustrates the causal nature of deformations in the RT curves as a result of quench

\begin{figure}[H]
\begin{center}
\includegraphics[scale=0.4]{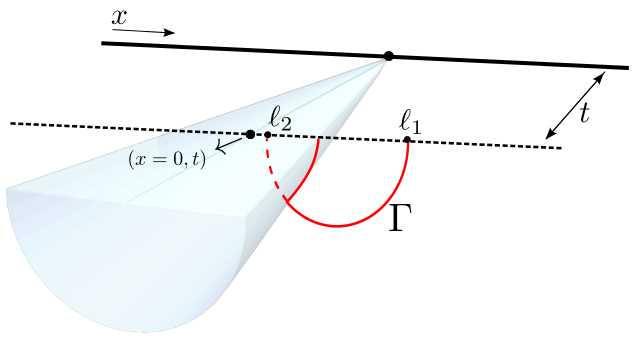}
\end{center}
\caption{Deformed RT curve.}
\label{fig9}
\end{figure}

The $t$-circle in this figure is the light front at time $t$.
This picture motivates us to propose that the light-cone of the falling
folded string deforms the RT curves as shown in figure(\ref{fig8}). We call the combination
in (\ref{decrease2}) as the {\it{decreasing\ contribution}} and denote it
by $S_D$. In figure (\ref{fig8}) we have introduced
two new curves, $\Gamma$ and $\tilde{\Gamma}$.
In figure(\ref{fig9}) we give a better illustration for the former.

In calculating the lengths of the new curves we come across
pieces that arrive at the boundary along the light front.
We should note that along such pieces the minimal value
for $z$, which appears as a limit in integrals, must be
set equal to $\delta$, the light-cone
regulator. Straightforward calculations lead to
\begin{equation}\label{deformed}
\CA(\Gamma)=\frac{c}{6}\ \log\frac{2t}{\delta}\
\frac{\ell_1-\ell_2}{a}\ \frac{\ell_1-t}{\ell_2+t}\ ,\ \ \ \\
\CA(\tilde{\Gamma})=\frac{c}{6}\ \log\frac{2t}{\delta}\
\frac{\ell_1+\ell_2}{a}\ \frac{\ell_1+t}{\ell_2+t}\ ,
\end{equation}
and the net contribution of the decreasing part, $S_D$, will be 
\begin{equation}
S_D=\CA(\Gamma)-\CA(\tilde{\Gamma})=\frac{c}{6}\
\log\frac{\ell_1-\ell_2}{\ell_1+\ell_2}\ \frac{\ell_1-t}{\ell_1+t}\ .
\end{equation}
As one can see, $a$ and $\delta$ dependence cancel out.
The situation is of course different once we
study the increasing contribution of $S_A$.

\begin{figure}[H]
\begin{center}
\includegraphics[scale=0.3]{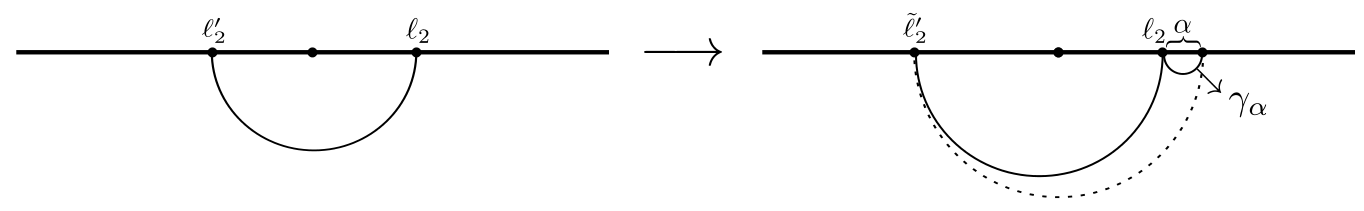}
\end{center}
\caption{Modifications to RT for the increasing contribution to entanglement.}
\label{fig10}
\end{figure}

Up to now we have addressed four of the initial RT curves and their possible modifications after quench.
We now turn to the remaining two. Of these, $\gamma_{\ell_1\ell'_1}$ is unaffected through the whole process
as it is outside the light cone. The other one however, $\gamma_{\ell_2\ell'_2}$, will  get modified 
in the increasing contribution to the EE.
This contribution, as stated before, is a result of the new entanglements
that the subset $\alpha$ finds with those degrees of freedom
that have already received the quench message. In figure (\ref{fig10}) we suggest
how to modify RT
for this contribution.

In the right hand side picture of figure (\ref{fig10}), where we have proposed
modifications to RT, the curve on the left is the deformation of the previously existing curve
$\gamma_{\ell_2\ell'_2}$. As the light front passes through
the image point $\ell'_2$, it sweeps away the image and drags the curve to
$\tilde{\ell}'_2$. 

We also propose that a new piece of curve, henceforth denoted by $\gamma_\alpha$,
forms between $\ell_2$ and the
light front on the right. This is to account for the new entanglements
that are formed between $\alpha$ and the interval $(\ell_2,\tilde{\ell}'_2)$.
Out of all the curves that we consider to find $S_A$,
this is the only one that is
produced during the process unlike the other ones which are deformations
of initial curves. In this sense the appearance of $\gamma_{\alpha}$ in our recipe is not as natural and satisfactory as 
the deformed curves we have proposed. In particular, and as we will see below,  we need to put a factor of $1/2$ 
in front of $\CA(\gamma_\alpha)$ to produce the correct results. This factor is a result of boundaries 
in the problem and we are aware that after quench when $\gamma_{\alpha}$ forms no boundaries exist.
It would be much more satisfactory if one could predict and motivate the formation of this new curve by 
field theory arguments. Unfortunately at the moment we do not have a compelling argument in favour of this part of our
proposal except that we have tested it in various cases when we change the relative position of the boundary and the interval and we have always 
produced the correct results. In particular we have applied the recipe to the cases of $\ell_1\rightarrow\infty, \ell_2=0$ and $\ell_2<0$
and found full agreement with field theory. We hope to report progress in this line in future works. 

In any case for the curves in figure (\ref{fig10}) one calculates
\begin{equation}\label{increase}
\CA(\gamma_{\ell_2\tilde{\ell}'_2})=\frac{c}{6}\log\frac{(t+\ell_2)^2}{a\delta}\ ,
\CA(\gamma_\alpha)=\frac{c}{6}\log\frac{(t-\ell_2)^2}{a\delta}\ ,
\end{equation}
which gives the increasing contribution, $S_I$ as
\begin{equation}\label{increase1}
S_I=\frac{1}{2}\CA(\gamma_{\ell_2\tilde{\ell}'_2})+\frac{1}{2}\CA(\gamma_\alpha)
=\frac{c}{6}\log\frac{t^2-\ell_2^2}{a\delta}\ .
\end{equation}
Putting everything together, we end up with
\begin{equation}\label{final}
S_A=S_D+S_I+\frac{1}{2}\CA(\gamma_{\ell_1\ell'_1})=\frac{c}{6}\ \log\frac{\ell_1-\ell_2}{\ell_1+\ell_2}\
\frac{\ell_1-t}{\ell_1+t}\ \frac{t^2-\ell_2^2}{a\delta}\ \frac{2\ell_1}{a}\ .
\end{equation}
Note that the last term is simply the only initial curve which has not
been affected by the light front. If we now express the holographic
light-cone regulator $\delta$ in terms of the field theory quench
regulator $\epsilon$ through equation (\ref{epsilondelta}), we recover
the field theory result of \cite{Calabrese:2006rx} in full agreement.

A last point worth mentioning is to see how the image points and their effects are totally erased
as the boundary is removed. This is accomplished by the deformed curves. 
By the time the quasi-particle reaches $\ell'_1$, that is when   $\tilde{\ell}'_2$ coincides with $\ell'_1$, $\CA(\gamma_{\ell'_1\tilde{\ell}'_2})$
tends to zero while the negative part due to $\CA(\gamma_{\ell_1\tilde{\ell}'_2})$ will be canceled by 
$\CA(\gamma_{\ell_1\ell'_2})$ which is one of the initial RT curves before quench. In addition,  
$\CA(\gamma_{\ell_2\tilde{\ell}'_2})$ will cancel the negative part $\CA(\gamma_{\ell_2\ell'_1})$ which
had remained unchanged by the modifications in the increasing contribution.

\begin{figure}[H]
\begin{center}
\includegraphics[scale=0.35]{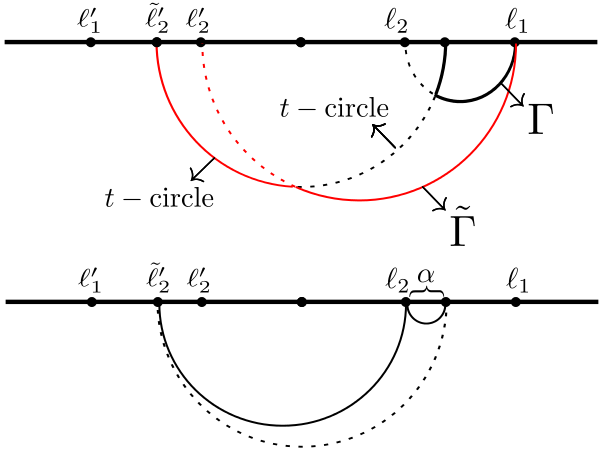}
\end{center}
\caption{Summary of modifications to RT; Above (below) produces $S_D$($S_I$).}
\label{fig11}
\end{figure}

In figure (\ref{fig11}) we summarize our proposal for the deformation and
modification of RT curves by the folded string light-cone.

\subsection{Interpretation of deformed RT curves}

The deformed curves $\Gamma$ and $\tilde{\Gamma}$ suggest a very
interesting interpretation. Recall the standard RT prescription
for holographic EE. A very illuminating interpretation of this
proposal was introduced in reference \cite{Casini:2011kv} where the EE of
a spherical region was explained in terms of the statistical entropy
of a related system in finite temperature. The latter system
has a gravitational dual which is a topological black hole.

As was shown in \cite{Casini:2011kv}, RT curve turns out to be the
horizon of the topological black hole and the bulk points which
are enclosed by this curve are the degrees of freedom outside the
horizon. This suggests that the reduced density matrix of the entangling
region can be described holographically by the bulk points enclosed by
the RT curve and that the length of the curve gives an account for the
number of states.

Now consider the RT curve $\gamma_{\ell_1\ell_2}$ which can be interpreted as above.
As the light-front of quench intersects this curve it deforms it into
$\Gamma$. This curve encloses only those points in the bulk which are
still {\it{unaware}} of quench. The length of this curve can be interpreted as how the original EE of the region is decreasing.
The new entanglements that are formed are accounted for by the points in
the bulk that are {\it{inside}} the light-cone. This, in principle, should be responsible for 
the new curve $\gamma_\alpha$.

Finally, one should note that there are other possibilities for the entangling interval. In all cases
our proposal for deformation of RT curves works perfectly and the results are
in complete agreement with field theory. 
%%%%%%%%%%%%%%%%%%%%%%%%%%%%%%%%%%%%%%%%%%%%%%%%%%%%%%%%%%%%%%%%%%%%%%%%%%%%%%%%%%%%%%%%%
%%%%%%%%%%%%%%%%%%%%%%%%%%%%%%%%%%%%%%%%%%%%%%%%%%%%%%%%%%%%%%%%%%%%%%%%%%%%%%%%%%%%%%%%%

\section{Discussion}
In this work we have proposed a holographic model for local quench
in $1+1$ dimensional CFT. Local quench is a result of joining two
identical BCFT's living on semi-infinite lines. We have assumed that
the initial theories have a zero boundary entropy.

Using AdS/BCFT our starting point in bulk is described
by two halves of AdS$_3$ which are bounded by the boundary
at $z=0$ on the one hand and a surface
$Q$ (and $\tilde{Q}$) on the other. We observe that for a
zero boundary entropy the surfaces $Q$ and $\tilde{Q}$
coincide with the tensionless Polyakov strings
which end at the boundary with a Neumann boundary condition.

Local quench is described holographically when the open strings
detach from boundary and form a closed string in the bulk. The
resulting closed string is a folded tensionless string whose
dynamics is given by the so called "Yo-Yo" solution. In particular
the tip of the folded string falls away from the boundary
on a light-like geodesic. Being tensionless, the string does not
back react on the geometry.

On the field theory side, the process of quench has a short
distance regulator which determines the locality of the
process. Also, the quasi-particles which are produced
by local manipulations propagate through the system and
produce the time dependence after quench. In our holographic
setup we introduce the geometric counterparts of these features.

The short distance regulator naturally appears as the minimal
distance between the tip of the folded string and the boundary.
We give an interpretation of this regulator in terms of that for
quench.
As the folding point travels on a light-like geodesic, a light-cone
propagates throughout the bulk. The origin of this cone is the space-time point
that the closed string has formed. The light-front is thus interpreted
as the bulk extension of the field theory quasi-particles.

We use the holographic model to calculate time dependent
entanglement entropy after quench. We propose how the light-cone
deforms standard RT curves and thus the EE. We find full agreement
with field theory results.

Dealing with a time dependent process, it is more natural to
apply the time dependent prescription to calculate holographic
EE according to \cite{Hubeny:2007xt}. Working in constant time slices
we avoid possible complications in this paper but it would be
nice to derive same results by the mentioned proposal.

A possibly interesting generalization of this work is to study the
same problem in the global coordinates. This would
correspond to a process of joining two CFT's on intervals
at the two ends and create a CFT on a circle. The bulk picture
will be a closed string that evolves in space.

A more challenging extension of this work is to study
local quench when the boundary entropy is nonzero. The
bulk surfaces $Q$ will not be tensionless in this case.
One should find the dynamics of the surfaces in the bulk and
should also take into account the back reaction on the geometry.
As pointed out in \cite{Calabrese:2006rx}, one expects that in a unitary
evolution of the system, the original excess of energy will not
dissipate and hence the system will always have a memory of the
initial condition. This feature was absent in the present
work as no energy excess existed in the first place.
This problem is currently under investigation.

\section*{Acknowledgements}

We would like to thank Mohsen Alishahiha for discussions. We would especially
like to thank Hessamaddin Arfaei for extensive discussions and also for a reading of the draft. We would like to thank the JHEP editor for a careful reading of the paper and for detailed questions and suggestions which has substantially improved our work and its presentation.

\section*{Appendix}
\appendix
\section{Geodesics in AdS$_3$}\label{appA}
In this appendix we will find geodesics in $AdS_3$ and calculate their length.\\

Consider the following parametrization for a geodesic, $\gamma_{x_1x_2}$, in AdS$_3$, \eqref{Poincare metric}
\begin{equation}
t=0\,\,\,\,\, , \,\,\,\,\, x_\gamma=x_\gamma(z)\, .
\end{equation}
which is a co-dimension two hypersurface whose induced metric reads
\begin{equation}
ds_{ind}^2=\frac{L^2}{z^2}\left(1+{x'_\gamma}^2\right)dz^2\, .
\end{equation}
Hence, one can evaluate the length of the curve as
\begin{equation}\label{area}
\CA=2\int dz \frac{L}{z}\sqrt{1+{x'_\gamma}^2}\, .
\end{equation}
The integrand can be considered as a Lagrangian which does not explicitly depend on variable $x_\gamma$
\begin{equation}
\CL=\CL[{x'_\gamma}(z)]\equiv \frac{L}{z}\sqrt{1+{x'_\gamma}^2}\, ,
\end{equation}
 therefore the equation of motion fixes the profile of the curve as
\begin{equation}
\frac{\p \CL}{\p {x'_\gamma}}=\text{const.}\rightarrow {x'_\gamma}(z)=\frac{z}{\sqrt{z_0^2-z^2}}\, ,
\end{equation}
where $z_0$ is the turning point at which ${x'_\gamma}\rightarrow\infty$.\\
Substituting this profile in \eqref{area} and performing the integral we arrive at
\begin{equation}
\CA=2\left(-L\log \frac{z_0+\sqrt{z_0^2-z^2}}{z}-L\log z_0\right)\Bigg\vert_a^{z_0}=2  L\log\left(\frac{2z_0}{a}\right) ,
\end{equation}
where $a$, represents the cut-off of the radial direction and $z_0=\frac{1}{2}\vert x_1-x_2\vert$.

%%%%%%%%%%%%%%%%%%%%%%%%%%%%%%%%%%%%%%%%%%%%%%%%%%%%%%%%%%%%%%%%%%%%%%%%%%%%%%%%%%%%%%

%%%%%%%%%%%%%%%%%%%%%%%%%%%%%%%%%%%%%%%%

\newpage
\begin{thebibliography}{99}
%\cite{cc0711}
\bibitem{cc0711}
P.Calabrese,J. Cardy
``Time Dependence of Correlation Functions Following a Quantum Quench"
Physical Review Letters, vol. 96, Issue 13, id. 136801
arXiv:cond-mat/0601225


%\cite{Calabrese:2007rg}
\bibitem{Calabrese:2007rg}
  P.~Calabrese and J.~Cardy,
  ``Quantum Quenches in Extended Systems,''
  J.\ Stat.\ Mech.\  {\bf 0706}, P06008 (2007)
  [arXiv:0704.1880 [cond-mat.stat-mech]].
  %%CITATION = ARXIV:0704.1880;%%
  %36 citations counted in INSPIRE as of 19 May 2014

  \bibitem{Calabrese:1002}
  P.~Calabrese and J.~Cardy,
  ``Quantum quench in interacting field theory: A Self-consistent approximation,''  Phys.\ Rev.\ B {\bf 81} (2010) 134305  [arXiv:1002.0167 [quant-ph]].  %%CITATION = ARXIV:1002.0167;%%

  %\cite{Maldacena:1997re},
\bibitem{Maldacena:1997re}
  J.~M.~Maldacena,
  ``The Large N limit of superconformal field theories and supergravity,''
  Adv.\ Theor.\ Math.\ Phys.\  {\bf 2}, 231 (1998)
  [hep-th/9711200].
  %%CITATION = HEP-TH/9711200;%%
  %9218 citations counted in INSPIRE as of 05 Sep 2013

%\cite{Witten:1998qj}
\bibitem{Witten:1998qj}
  E.~Witten,
  ``Anti-de Sitter space and holography,''
  Adv.\ Theor.\ Math.\ Phys.\  {\bf 2}, 253 (1998)
  [hep-th/9802150].
  %%CITATION = HEP-TH/9802150;%%
  %6187 citations counted in INSPIRE as of 05 Sep 2013

\bibitem{Aharony}
  O.~Aharony, S.~S.~Gubser, J.~M.~Maldacena, H.~Ooguri and Y.~Oz,
  ``Large N field theories, string theory and gravity,''  Phys.\ Rept.\  {\bf 323} (2000) 183  [hep-th/9905111].  %%CITATION = HEP-TH/9905111;%%


%\cite{Calabrese:2006rx}
\bibitem{Calabrese:2006rx}
  P.~Calabrese and J.~L.~Cardy,
 ``Entanglement and correlation functions following a local quench: a conformal field theory approach,``
  J. Stat. Mech. P10004(2007) P10004
  [cond-mat/0708.3750].
%\cite{}
\bibitem{2007JSMTE..06....5E}
   Eisler, V. and Peschel, I.,
   `` Evolution of entanglement after a local quench,``
   J.\ Stat.\ Mech.\ (2007) P06005, cond-mat/0703379;

\bibitem{2008JSMTE..01..023E},
   {Eisler}, V. and {Karevski}, D. and {Platini}, T. and {Peschel}, I.
	,
    ``Entanglement evolution after connecting finite to infinite quantum chains``,
J.\ Stat.\ Mech.\ (2008) P01023, arXiv:0711.0289.

\bibitem{2011JSMTE..10..027D}
   {{Divakaran}, U. and {Igl{\'o}i}, F. and {Rieger}, H.},
    ``Non-equilibrium quantum dynamics after local quenches,
  {Journal of Statistical Mechanics: Theory and Experiment},

      doi = {10.1088/1742-5468/2011/10/P10027},




%\cite{Calabrese:2009qy}
\bibitem{Calabrese:2009qy}
  P.~Calabrese and J.~Cardy,
  ``Entanglement entropy and conformal field theory,''
  J.\ Phys.\ A {\bf 42}, 504005 (2009)
  [arXiv:0905.4013 [cond-mat.stat-mech]].
  %%CITATION = ARXIV:0905.4013;%%
  %126 citations counted in INSPIRE as of 19 May 2014
 %\cite{Aparicio:2011zy},

%\cite{Hartman:2013qma}
\bibitem{Hartman:2013qma}
  T.~Hartman and J.~Maldacena,
  ``Time Evolution of Entanglement Entropy from Black Hole Interiors,''
  JHEP {\bf 1305}, 014 (2013)
  [arXiv:1303.1080 [hep-th]].
  %%CITATION = ARXIV:1303.1080;%%
  %12 citations counted in INSPIRE as of 16 Jul 2013


\bibitem{Aparicio:2011zy}
  J.~Aparicio and E.~Lopez,
  ``Evolution of Two-Point Functions from Holography,''
  JHEP {\bf 1112}, 082 (2011)
  [arXiv:1109.3571 [hep-th]].
  %%CITATION = ARXIV:1109.3571;%%
  %28 citations counted in INSPIRE as of 05 Sep 2013

%\cite{Albash:2010mv}
\bibitem{Albash:2010mv}
  T.~Albash and C.~V.~Johnson,
  ``Evolution of Holographic Entanglement Entropy after Thermal and Electromagnetic Quenches,''
  New J.\ Phys.\  {\bf 13}, 045017 (2011)
  [arXiv:1008.3027 [hep-th]].
  %%CITATION = ARXIV:1008.3027;%%
  %51 citations counted in INSPIRE as of 05 Sep 2013

%\cite{Takayanagi:2011zk}
\bibitem{Takayanagi:2011zk}
  T.~Takayanagi,
  ``Holographic Dual of BCFT,''
  Phys.\ Rev.\ Lett.\  {\bf 107}, 101602 (2011)
  [arXiv:1105.5165 [hep-th]].
  %%CITATION = ARXIV:1105.5165;%%
  %42 citations counted in INSPIRE as of 19 May 2014

%\cite{Fujita:2011fp}
\bibitem{Fujita:2011fp}
  M.~Fujita, T.~Takayanagi and E.~Tonni,
  ``Aspects of AdS/BCFT,''
  JHEP {\bf 1111}, 043 (2011)
  [arXiv:1108.5152 [hep-th]].
  %%CITATION = ARXIV:1108.5152;%%
  %32 citations counted in INSPIRE as of 19 May 2014

  %\cite{Bardeen:1975gx}
\bibitem{Bardeen:1975gx}
  W.~A.~Bardeen, I.~Bars, A.~J.~Hanson and R.~D.~Peccei,
  ``A Study of the Longitudinal Kink Modes of the String,''
  Phys.\ Rev.\ D {\bf 13}, 2364 (1976).
  %%CITATION = PHRVA,D13,2364;%%
  %124 citations counted in INSPIRE as of 19 May 2014

  %\cite{Ficnar:2013wba}
\bibitem{Ficnar:2013wba}
  A.~Ficnar and S.~S.~Gubser,
  ``Finite momentum at string endpoints,''
  Phys.\ Rev.\ D {\bf 89}, 026002 (2014)
  [arXiv:1306.6648 [hep-th]].
  %%CITATION = ARXIV:1306.6648;%%
  %5 citations counted in INSPIRE as of 19 May 2014

%\cite{Gubser:2002tv}
\bibitem{Gubser:2002tv} 
  S.~S.~Gubser, I.~R.~Klebanov and A.~M.~Polyakov,
  ``A Semiclassical limit of the gauge / string correspondence,''
  Nucl.\ Phys.\ B {\bf 636}, 99 (2002)
  [hep-th/0204051].
  %%CITATION = HEP-TH/0204051;%%
  %793 citations counted in INSPIRE as of 27 Mar 2015

  %\cite{Ryu:2006bv}
\bibitem{Ryu:2006bv}
  S.~Ryu and T.~Takayanagi,
  ``Holographic derivation of entanglement entropy from AdS/CFT,''
  Phys.\ Rev.\ Lett.\  {\bf 96}, 181602 (2006)
  [hep-th/0603001].
  %%CITATION = HEP-TH/0603001;%%
  %466 citations counted in INSPIRE as of 19 May 2014

  %\cite{Nishioka:2009un}
\bibitem{Nishioka:2009un}
  T.~Nishioka, S.~Ryu and T.~Takayanagi,
  ``Holographic Entanglement Entropy: An Overview,''
  J.\ Phys.\ A {\bf 42}, 504008 (2009)
  [arXiv:0905.0932 [hep-th]].
  %%CITATION = ARXIV:0905.0932;%%
  %199 citations counted in INSPIRE as of 19 May 2014

  %\cite{Ryu:2006ef}
\bibitem{Ryu:2006ef}
  S.~Ryu and T.~Takayanagi,
  ``Aspects of Holographic Entanglement Entropy,''
  JHEP {\bf 0608}, 045 (2006)
  [hep-th/0605073].
  %%CITATION = HEP-TH/0605073;%%
  %339 citations counted in INSPIRE as of 19 May 2014


  %\cite{Nozaki:2013wia}
\bibitem{Nozaki:2013wia}
  M.~Nozaki, T.~Numasawa and T.~Takayanagi,
  ``Holographic Local Quenches and Entanglement Density,''
  JHEP {\bf 1305}, 080 (2013)
  [arXiv:1302.5703 [hep-th]].
  %%CITATION = ARXIV:1302.5703;%%
  %28 citations counted in INSPIRE as of 19 May 2014


   %\cite{Ugajin:2013xxa}
\bibitem{Ugajin:2013xxa}
  T.~Ugajin,
  ``Two dimensional quantum quenches and holography,''
  arXiv:1311.2562 [hep-th].
  %%CITATION = ARXIV:1311.2562;%%
  %3 citations counted in INSPIRE as of 19 May 2014

%\cite{Asplund:2013zba}
\bibitem{Asplund:2013zba}
  C.~T.~Asplund and A.~Bernamonti,
  ``Mutual information after a local quench in conformal field theory,''
  Phys.\ Rev.\ D {\bf 89}, 066015 (2014)
  [arXiv:1311.4173 [hep-th]].
  %%CITATION = ARXIV:1311.4173;%%
  %1 citations counted in INSPIRE as of 19 May 2014

%\cite{Basu:2011ft}
\bibitem{Basu:2011ft}
  P.~Basu and S.~R.~Das,
  ``Quantum Quench across a Holographic Critical Point,''
  JHEP {\bf 1201}, 103 (2012)
  [arXiv:1109.3909 [hep-th]].
  %%CITATION = ARXIV:1109.3909;%%
  %27 citations counted in INSPIRE as of 20 May 2014

  %\cite{Das:2011nk}
\bibitem{Das:2011nk}
  S.~R.~Das,
  ``Holographic Quantum Quench,''
  J.\ Phys.\ Conf.\ Ser.\  {\bf 343}, 012027 (2012)
  [arXiv:1111.7275 [hep-th]].
  %%CITATION = ARXIV:1111.7275;%%
  %15 citations counted in INSPIRE as of 20 May 2014



  %\cite{Buchel:2012gw}
\bibitem{Buchel:2012gw}
  A.~Buchel, L.~Lehner and R.~C.~Myers,
  ``Thermal quenches in N=2* plasmas,''
  JHEP {\bf 1208}, 049 (2012)
  [arXiv:1206.6785 [hep-th]].
  %%CITATION = ARXIV:1206.6785;%%
  %25 citations counted in INSPIRE as of 20 May 2014


  %\cite{Basu:2012gg}
\bibitem{Basu:2012gg}
  P.~Basu, D.~Das, S.~R.~Das and T.~Nishioka,
  ``Quantum Quench Across a Zero Temperature Holographic Superfluid Transition,''
  JHEP {\bf 1303}, 146 (2013)
  [arXiv:1211.7076 [hep-th]].
  %%CITATION = ARXIV:1211.7076;%%
  %17 citations counted in INSPIRE as of 20 May 2014

  %\cite{Buchel:2013lla}
\bibitem{Buchel:2013lla}
  A.~Buchel, L.~Lehner, R.~C.~Myers and A.~van Niekerk,
  ``Quantum quenches of holographic plasmas,''
  JHEP {\bf 1305}, 067 (2013)
  [arXiv:1302.2924 [hep-th]].
  %%CITATION = ARXIV:1302.2924;%%
  %21 citations counted in INSPIRE as of 20 May 2014


%\cite{Pedraza:2014moa}
\bibitem{Pedraza:2014moa} 
  J.~F.~Pedraza,
  ``Evolution of nonlocal observables in an expanding boost-invariant plasma,''
  Phys.\ Rev.\ D {\bf 90}, 046010 (2014)
  [arXiv:1405.1724 [hep-th]].
  %%CITATION = ARXIV:1405.1724;%%
  %6 citations counted in INSPIRE as of 27 Nov 2014

  %\cite{Roberts:2012aq}
\bibitem{Roberts:2012aq}
  M.~M.~Roberts,
  ``Time evolution of entanglement entropy from a pulse,''
  arXiv:1204.1982 [hep-th].
  %%CITATION = ARXIV:1204.1982;%%
  %4 citations counted in INSPIRE as of 20 Jul 2013

%\cite{DiFrancesco:1997nk}
\bibitem{DiFrancesco:1997nk} 
  P.~Di Francesco, P.~Mathieu and D.~Senechal,
  ``Conformal field theory,''
  New York, USA: Springer (1997) 890 p
  %42 citations counted in INSPIRE as of 28 Dec 2014
  
  %\cite{Hubeny:2007re}
\bibitem{Hubeny:2007re} 
  V.~E.~Hubeny and M.~Rangamani,
  ``Holographic entanglement entropy for disconnected regions,''
  JHEP {\bf 0803}, 006 (2008)
  [arXiv:0711.4118 [hep-th]].
  %%CITATION = ARXIV:0711.4118;%%
  %38 citations counted in INSPIRE as of 03 Jan 2015


%\cite{Casini:2011kv}
\bibitem{Casini:2011kv}
  H.~Casini, M.~Huerta and R.~C.~Myers,
  ``Towards a derivation of holographic entanglement entropy,''
  JHEP {\bf 1105}, 036 (2011)
  [arXiv:1102.0440 [hep-th]].
  %%CITATION = ARXIV:1102.0440;%%
  %181 citations counted in INSPIRE as of 20 May 2014

%\cite{Hubeny:2007xt}
\bibitem{Hubeny:2007xt}
  V.~E.~Hubeny, M.~Rangamani and T.~Takayanagi,
  ``A Covariant holographic entanglement entropy proposal,''
  JHEP {\bf 0707}, 062 (2007)
  [arXiv:0705.0016 [hep-th]].
  %%CITATION = ARXIV:0705.0016;%%
  %174 citations counted in INSPIRE as of 20 May 2014

\end{thebibliography}
\end{document}